\documentclass[english,prd,twocolumn,superscriptaddress,nofootinbib,preprintnumbers]{revtex4}
\usepackage[latin1]{inputenc}
\usepackage{graphicx}
\usepackage{amssymb}
\usepackage{color}
\usepackage{float}
\usepackage{amsmath}
\usepackage{amsfonts}
\usepackage{dcolumn}
\usepackage{hyperref}
\usepackage{subfigure}

\begin{document}
 
\title{Neutron stars in Starobinsky model}

\author{Apratim Ganguly}
\author{Radouane Gannouji}
\author{Rituparno Goswami}
\author{Subharthi Ray}
\affiliation{Astrophysics and Cosmology Research Unit, School of Mathematics, Statistics and Computer Sciences, University of KwaZulu-Natal, Private Bag X54001, Durban 4000, South Africa}


\begin{abstract}
We study the structure of neutron stars in $f(R)=R+\alpha R^{2}$ theory of gravity (Starobinsky model), in an exact and non-perturbative approach. In this model, apart from the standard General Relativistic junction conditions, two extra conditions, namely the continuity of the curvature scalar and its first derivative needs to be satisfied. For an exterior Schwarzschild solution, the curvature scalar and its derivative has to be zero at the stellar surface. We show that for some equation of state (EoS) of matter, matching all conditions at the surface of the star is impossible. Hence the model brings two major fine-tuning problems: (i) only some particular classes of EoS are consistent with Schwarzschild at the surface and (ii) given that EoS, only a very particular set of boundary conditions at the centre of the star will satisfy the given boundary conditions at the surface. Hence we show that this model (and subsequently many other $f(R)$ models where uniqueness theorem is valid) is highly unnatural, for the existence of compact astrophysical objects. This is because the EoS of a compact star should be completely determined by the physics of nuclear matter at high density and not the theory of gravity.
\end{abstract}


\maketitle
 
\section{Introduction}
\label{intro}

In order to solve the problems (viz. flatness, horizon, monopoles) concerning the early universe in cosmology, a phase of very rapid accelerated expansion is necessary \cite{Guth:1980zm,Linde:1981mu,Albrecht:1982wi}, the inflation. Furthermore, it naturally provides an initial seed for CMB anisotropy and large scale structure \cite{Mukhanov:1981xt,Hawking:1982cz,Starobinsky:1982ee,Guth:1982ec}. Hence it became a cornerstone of the Big Bang model. Many models to explain this early acceleration are in the market. The most popular models contain a scalar field with a slowly varying potential. There is another class of models in which gravity
is modified with respect to General Relativity (GR). One of the schemes of these modifications based upon phenomenological considerations is provided by $f(R)$ theories of gravity. These theories essentially contain an additional scalar degree of freedom apart from graviton. Indeed, $f(R)$ theories are conformally equivalent to Einstein theory plus a canonical scalar degree of freedom dubbed scalaron whose potential is uniquely constructed from Ricci scalar. One of the main interesting aspect of these models of gravity (contrary to models including Ricci or Riemann tensor in the action) is the absence of Ostrogradski ghost  despite the equations of motion being of fourth order.\\
In literature, there are various models where the authors \cite{Ginzburg,Bunch,Davies} considered Einstein equations with quantum corrections. Following these ideas, Starobinsky studied the cosmology of one of these models \cite{Starobinsky:1980te} which later has been simplified and popularly known today as the Starobinsky's model, where the action of General Relativity is replaced by $f(R)=R+\alpha R^2$. It has been the first internally consistent inflationary model. In this model, the $R^2$ term produces an accelerated stage in the early universe preceding the usual radiation and matter stages. Notice that contrary to the Starobinsky model, generic $f(R)$ models are plagued by various problems; they generally reduce to GR plus cosmological constant \cite{Thongkool:2009js} or have a $\phi$MDE (field-matter dominated epoch) instead of a standard matter epoch \cite{Amendola:2006we} or hit curvature singularity \cite{Frolov:2008uf} (see \cite{Kobayashi:2008tq} for the existence of a singularity in an asymptotic de Sitter Universe and \cite{Upadhye:2009kt} for the opposite statement), or produce high frequency oscillations and a singularity at finite time in cosmology \cite{Appleby:2008tv} or give rise a fine tuning \cite{Faraoni:2011pm}. \\
The recent results from the Planck satellite \cite{Ade:2013uln} are remarkably compatible with the Starobinsky's model. Hence the model remains one of the candidate of gravity at high energies in the early epoch of the Universe and avoids the difficulties listed previously. \\
A consistent theory of gravity, classical or modified, should be equally applicable to the strong gravity regime. As a common practice, all the modified gravity models are verified of their feasibility by undergoing the solar system test. However, these fields are substantially weaker than  the vicinity of the astrophysical compact stars and solar mass black holes corresponding to a surface redshift of $\sim$ 1 and a spacetime curvature of $\simeq 2 \times 10^{-13} {\rm cm}^{-2}$  \cite{Psaltis:2008bb}. So these compact stars and solar mass black holes give a very good platform to study the behaviour of strong gravity. Ideally, these black holes would have been the best candidates to study the strong gravity behaviour of these modified theories. On the other hand, the compact stars like the neutron and quark stars have additional benefits of studying behaviour of matter at high density under the modification of gravity. However, verification and constraining the EoS of matter at high density, with the aid of these modified gravity theories, if any, come at a much later stage. The challenges we face here is to find a proper matching of the exterior solution with the interior. A few attempts to apply the modified gravity models to neutron stars has been done in the recent past (see e.g.~\cite{Kainulainen:2007bt,Babichev:2009td,Cooney:2009rr,Babichev:2009fi,Arapoglu:2010rz,Orellana:2013gn,Alavirad:2013paa,Astashenok:2013vza}). But as far as we know, there is no work where the interior solution has been consistently matched with a viable exterior spacetime. As we will see, this will drastically change the conclusions on the viability of the model.\\
The paper is organized as follows. We first, very briefly review the basic equations (TOV) of the model in section \ref{formalisme}, followed by the junction conditions at the surface of the star in section \ref{junc}. In the section \ref{Schwar} we motivate why Schwarzschild solution should be the exterior solution. In section \ref{contraintes} we give the laboratory, solar system and cosmological constraints on the parameter $\alpha$ of the model. The section \ref{singular} is devoted to the singular problem of the system with the existence of boundary layers at the surface. In sections \ref{0directe} and \ref{directe}, we perform a numerical analysis to confirm the fine-tuned nature of the problem and hence the difficulty to match all boundary conditions. Finally, in the section \ref{pasdirecte} we propose a semi analytical approach to solve the problem. The last section is devoted to summary and discussion.

\section{Action and Modified TOV equations}
\label{formalisme}

The straightforward generalisation of the Lagrangian in the Einstein-Hilbert action results in 4-dimensional action in $f(R)$ gravity :

\begin{align}
\label{b1}
\mathcal{S} = \frac{1}{2\kappa^{2}}\int d^{4}x\sqrt{-g}f(R)+ \mathcal{S}_{m}\,,
\end{align}
where $g$ denotes the determinant of the metric $g_{\mu \nu}$ and $R$ is the Ricci scalar. $G$ and $c$ are set to 1 in the rest of the paper, hence $\kappa^{2}=8 \pi$ and $S_{m}$ is the action for matter.
 
Following the metric formalism from the above action, the field equations are derived by the variation of the action with respect to the metric tensor $g_{\mu \nu}$, and we get

\begin{align}
  \label{b2}
  F(R)R_{\mu\nu}-\frac{1}{2}\emph{f(R)}g_{\mu\nu}-(\nabla_{\mu}\nabla_{\nu}-g_{\mu\nu}\Box)F(R)=\kappa^{2}T_{\mu\nu}\,,
 \end{align}
where $F(R)=\frac{df(R)}{dR}$ and $T_{\mu \nu}=-\frac{2}{\sqrt{-g}}\frac{\delta S_{m}}{\delta g^{\mu \nu}}$ is the energy-momentum tensor of the matter fields. \\
Substituting $f(R)=R+\alpha R^{2}$, we get

 \begin{align}
  \label{b45}
&G_{\mu\nu}(1+2\alpha R) +\frac{\alpha}{2} g_{\mu\nu}R^{2}-2\alpha(\nabla_{\mu}\nabla_{\nu}-g_{\mu\nu}\Box)R =\kappa^{2}T_{\mu\nu}\,, \\
&6\alpha \Box R-R =\kappa^{2}T\,,
 \end{align}
where $G_{\mu \nu}=R_{\mu \nu}-\frac{1}{2}g_{\mu \nu}R$ is the Einstein tensor.\\
 
As we are interested in spherically symmetric solutions of these field equations inside a neutron star, we choose a spherically symmetric metric of the form

\begin{align}
\label{b6}
ds^{2}=-e^{2\Phi(r)}dt^{2}+\left(1-\frac{2m(r)}{r}\right)^{-1}dr^{2}+r^{2}d\Omega^2\,.
\end{align}
Using the metric, the (0,0) and (1,1) component, the trace of the Einstein's equations and the conservation equation, we obtain the following system of equations

\begin{widetext}
  \begin{align}
  \label{b9:1}
  m' &= \frac{1}{12(1+2\alpha R)(1+2\alpha R+\alpha rR')} \left[r^{2}(1+2\alpha R)(48 \pi P+R (2+3\alpha R)+32 \pi \rho)\right. \nonumber \\
  &\qquad \qquad \qquad \qquad\left. +2\alpha (-6m(1+2\alpha R)+r^{3}(R+3\alpha R^{2}+16\pi \rho))R'+24\alpha^{2} r(r-2m)R'^{2}\right]\,,\\
  \label{b9:2}
  P' &= -\frac{(P+ \rho)(4m+16\pi r^{3}P+8\alpha mR-\alpha r^{3}R^{2}-8\alpha r(r-2m)R')}{4r(r-2m)(1+2\alpha R+\alpha rR')}\,,\\
  \label{b9:3}
  R'' &= \frac{1}{6\alpha r(r-2m)(1+2\alpha R)}\left[r^{2}(1+2\alpha R)(24\pi P+R-8\pi \rho)+\alpha (12m(1+2\alpha R) \right. \nonumber \\
  & \qquad \qquad \qquad\qquad \left. +r(-12+R(r^{2}-24\alpha+3\alpha r^{2} R)+16\pi r^{2} \rho))R'+12\alpha^{2}r(r-2m)R'^{2}\right]\,,
  \end{align}
\end{widetext}

where prime denotes derivative with respect to radial distance $(r)$. Finally, an EoS $P=P(\rho)$ closes the set of equations (\ref{b9:1},\ref{b9:2},\ref{b9:3}).

\section{Junction Conditions}
\label{junc}
In what follows we will be matching together the interior of the star to a well defined exterior geometry,  in order to construct a realistic neutron star model.  This requires a set of junction conditions, analogous to the Israel junction conditions from general relativity ~\cite{Israel:1966rt}, and is a problem that has been considered in $f(R)$ theories of gravity by Deruelle, Sasaki and Sendouda ~\cite{Deruelle:2007pt} and later by other authors ~\cite{clifton,senovilla}.  We will briefly recap the relevant results from their work here, as it of central importance to our study.\\
The prime requirement in ~\cite{Deruelle:2007pt} is that if one allows delta functions on the matter part of the field equations (i.e. if one allows matter fields to be localised on the boundary hyper-surface), then delta functions should occur at most linearly in the parts of the field equations that involve geometry only. Here we are interested in the case in which there is no brane located at the boundary. We therefore require that there should be {\it no} delta function in the part of the field equations containing just the geometry.
Therefore, in $f(R)$ gravity theories, apart from the usual GR junction conditions i.e. the agreement of the first and second fundamental forms on the both sides of the matching timelike hypersurface,
 \begin{align}
 \label{c1}
 [h_{\mu\nu}]=0, \qquad [K_{\mu\nu}]=0\,,
 \end{align} 
where $h_{\mu\nu}$ and $K_{\mu\nu}$ are the first and the second fundamental forms respectively and $[~]$ denotes the jump across the surface, two more conditions need to be satisfied. These are the continuity of the scalar curvature and its first derivative across the boundary
  \begin{align}
 \label{c2}
 [R]=0, \qquad [\nabla_{\mu}R]=0.
 \end{align}  
 These extra conditions above makes the problem of matching the stellar interior with a suitable exterior spacetime extremely restrictive. In fact most of the compact star models discussed so far in higher order theories of gravity, do not take into account these restrictions rigorously \cite{Kainulainen:2007bt,Babichev:2009td,Cooney:2009rr,Babichev:2009fi,Arapoglu:2010rz,Orellana:2013gn,Alavirad:2013paa,Astashenok:2013vza}.
 
\section{Exterior Spacetime: Why Schwarzschild?}
\label{Schwar}
We know in general relativity, Jebsen-Birkhoff's theorem states that the Schwarzschild solution is the unique spherically symmetric solution
of the vacuum Einstein field equations. In that case  a spherically symmetric gravitational field in empty space outside a spherical star must be static, with a metric given by the Schwarzschild metric (for $r>2M$)
\begin{align}
\label{Sch}
ds^2 = -\left(1-\frac{2M}{r}\right)dt^2 + \frac{dr^2}{\left(1-\frac{2M}{r}\right)}+r^2 d \Omega^2\;.
\end{align}
It represents the spacetime of the Solar System, and all other spherically symmetric stellar
systems, to a very good approximation, and hence forms the key geometry in much of astrophysics and astronomy. 

For higher order gravity theories, the Jebsen-Birkhoff theorem in its original form is violated. As we can easily see, in $f(R)$ theories, the trace equation in vacuum is a massive Klein-Gordon equation, which has different classes of non-trivial exact vacuum solutions that can be both static and non-static (see for example \cite{clifton1}). Hence, in principle, there exists  a larger freedom for the exterior spacetime outside a static star. However, since we are interested in modelling  realistic astrophysical compact objects, we will require the exterior spacetime to be static and asymptotically flat, as dictated by observational tests within the solar system. Furthermore, whatever exterior static solution we use for 
a compact star, the same should describe a well defined Black hole solution, as the astrophysical Black holes are formed via gravitational collapse of these compact objects. Given the above constraints, there are two possible ways to construct a suitable exact exterior for the astrophysical compact stars:
\begin{enumerate}
\item {\bf  Matching the interior with an exact asymptotically flat static solution}. Though a very few exact static vacuum solutions are known for the Starobinsky model of $R+\alpha R^2$ gravity, there exists the following uniqueness theorem \cite{Whitt:1984pd,Mignemi:1991wa} : {\it For all functions $f(R)$ which are of class $C^3$ at $R=0$ and $f(0)=0$ while $f'(0)\ne 0$, the only static spherically symmetric asymptotically flat solution with a regular horizon in these models is the Schwarzschild solution, provided that the coefficient of the $R^2$ term in the Lagrangian polynomial is positive}. Since we require $\alpha>0$ to avoid ghosts in the theory and also require the solution to describe a well defined Black Hole with a regular horizon, Schwarzschild solution is the only possible exact asymptotically flat exterior. This is a very well known result which follows the famous BH no-scalar-hair theorems. It states that stationary BH solution are the same as those in general relativity, namely Schwarzschild for the non-rotating case. It was proved by Bekenstein \cite{Bekenstein:1995un} and Sudarsky \cite{Sudarsky:1995zg} for a quintessence field with convex potential, which corresponds to Starobinsky model in the Einstein frame. Also an extension has been established without assuming any symmetries apart from stationarity in \cite{Sotiriou:2011dz}. Obviously if one relaxes the condition of asymptotic flatness and allows for asymptotically de Sitter spacetime, one might have other solutions, e.g. by adding a cosmological constant to the model. But the BH no-scalar-hair theorem has been extended to this case in \cite{Torii:1998ir}. Therefore static BH asymptotically flat or de-Sitter are no different than in general relativity. Finally we might think to a non-static solution, so far all the approximate solutions derived in the literature develop high time oscillating modes because of the presence of the scalaron and should be discarded (see e.g. \cite{Starobinsky:2007hu} in the cosmological case).
\item{\bf Matching the interior to an intermediate static vacuum solution that can be matched to Schwarzschild at a larger distance}. Let us consider an intermediate 
non-Schwarzschild static exterior matched with the stellar boundary $r=r_s$ and this intermediate solution is then matched to Schwarzschild at $r=r_2$, where $r_2>r_s$. On a superficial level this construction seems to have more freedom as the intermediate static solution need not be asymptotically flat. However, keeping in mind the physicality conditions, we would like this intermediate region to be smooth ($C^\infty$). Now from the matching conditions one can immediately see that the Ricci Scalar ($R$) and its normal derivative ($R'$) should vanish at the outer boundary $r=r_2$ of this intermediate solution, where we are matching with the Schwarzschild spacetime. Using the trace equation (\ref{b9:3}), we get $R''(r_2)=0$ and then the smoothness implies that all the subsequent derivatives of the Ricci scalar must vanish at this boundary. Therefore there exists an open neighbourhood 
$\mathcal{U}\ni r_2$ where $R=0$. By continuity, we can then extend this open neighbourhood to the entire exterior asymptotically flat submanifold. Therefore the Ricci scalar vanishes at every point on the exterior submanifold. Using the extension of Jebsen-Birkhoff theorem to $f(R)$ gravity ~\cite{Nzioki:2009av}, which states that {\it for all functions $f(R)$ which are of class $C^3$ at $R=0$ and $f(0)=0$ while $f'(0)\ne 0$, Schwarzschild solution is the only vacuum spherically symmetric solution with vanishing Ricci scalar}, we can immediately show that this intermediate region has to be Schwarzschild. We also note that this proof remains true even if we consider more than one intermediate regions between the stellar boundary and Schwarzschild.
\end{enumerate}
It therefore seems natural to match the spherically symmetric static star with a Schwarzschild exterior. However the crucial difference with general relativity comes from the fact that for this case, the matching surface, the Ricci scalar and it's normal derivative must vanish. This makes the interior solution much more restrictive than GR.

\section{Constraints on the model}
\label{contraintes}
In this section, we derive the various constraints  on the parameter $\alpha$. Let us first consider the experimental bound that comes from the solar system tests of the equivalence principle (LLR). For any chameleon theory with a scalar field $(\phi)$ we can define a thin-shell parameter $\epsilon$ \cite{Khoury:2003aq}, which for the Earth gives

\begin{align}
\epsilon\equiv \sqrt{6}\frac{\phi_{\infty}-\phi_{\oplus}}{M_{pl}\Phi_\oplus}<2.2\times10^{-6}
\end{align}
where $(\phi_\infty,\phi_\oplus)$ are respectively the minimum of the
effective potential at infinity and inside the planet and $\Phi_\oplus$ the Newton potential for the Earth.
Notice that the constraint on the post-Newtonian parameter  $\gamma$ gives $\epsilon<2.3\times 10^{-5}$.
Using the value $\Phi_\oplus\simeq 7\times 10^{-10}$, the previous bound
translates into $\phi_\infty/M_{pl}<10^{-15}$.\\
We know that any $f(R)$-theory can be written into a chameleon form after a conformal transformation, the 2 frames are physically equivalent. Hence the model can be cast in the form of a scalar field in an effective potential. The existence of the chameleon mechanism depends on the form of the effective potential which in turn depends on the local density and pressure. When pressure is negligible and density is large, the scalar field may acquire
a large mass for a suitably chosen potential leading to suppression of the fifth force locally. 
The scalar field is assumed to be settled in the minimum of the effective potential.
Hence it is easy to find that the minimum of the effective potential can be written in the Jordan frame in the following form
\begin{align}
2f(R)-R f'(R)=\frac{\rho-3 P}{M_{pl}^2}\,,
\end{align}
which corresponds to the trace equation in the constant curvature case. It turns out that the $\alpha R^2$ term do not change the minimum of the effective potential. Hence the LLR bound leads to \cite{Gannouji:2012iy}

\begin{align}
\label{coucou}
\Bigl|f'\Bigl(\frac{\rho_\infty}{M_{pl}^2}\Bigr)-1\Bigr|<10^{-15}.
\end{align}
For the Starobinsky model and with the density $\rho_\infty \simeq 10^{-24} \text{g cm}^{-3}$, eq.(\ref{coucou}) tells us that $\alpha<10^{-15}M_{pl}^2/\rho_m$ which gives $\alpha<10^{45}\text{eV}^{-2}$.\\
But the tightest local constraint comes from the E{\"o}t-Wash experiments, which use torsion balances.
We know that a point mass has a Yukawa gravitational potential (see e.g. \cite{Stelle:1977ry})
\begin{align}
V(r)=\frac{GM}{r}\Bigl(1+\frac{1}{3}e^{-r/\sqrt{6\alpha}}\Bigr)\,,
\end{align}
which gives \cite{Kapner:2006si} $\alpha<4\times 10^4 \text{eV}^{-2}$. Notice that according to the bound from Big Bang nucleosynthesis and CMB physics, we have $\alpha\ll 10^{35} \text{eV}^{-2}$ \cite{Zhang:2007ne}. \\
We turn now to the inflation, and according to the latest dataset from Planck, the Starobinsky model is a viable candidate for the early acceleration phase of the Universe. We have  \cite{Starobinsky:2007hu, Starobinsky:1983zz} $\alpha\simeq 10^{-45}\bigl(N/50\Bigr)^2 \text{eV}^{-2}$ where $N$ is the number of e-folds. Notice that it may not be compatible with the classicality condition of the field \cite{Gannouji:2012iy,Upadhye:2012vh}.

Hence we conclude this section by considering $\alpha\simeq 10^{-45} \text{eV}^{-2}$ from the cosmological constraints or $\alpha<4\times 10^4 \text{eV}^{-2}$ from the laboratory tests.

\section{Singular problem}
\label{singular}

As it was noticed in various papers, the equations (\ref{b9:1},\ref{b9:2},\ref{b9:3}) are very complicated to solve for realistic cases where $\alpha$ is very small. Also, often in the literature, a simple series expansion is used to carry out these calculations. But as it is known, the solution in powers of the natural small parameter $(\alpha\ll 1)$ is invalid if we are in the case of a boundary layer problem also known as a singular perturbation (see e.g. \cite{Nayfeh}). In fact, our equations are among these last problems because a small parameter $(\alpha)$ multiplies the highest order derivative, and we have $\alpha R''+\cdots$ in \eqref{b9:3}. We should also mention that it is not a priori clear that a non-linear boundary value problem has a solution. \\
A singular problem is associated with approximation of \eqref{b9:3} for small values of $\alpha$. The difficulty near the boundaries arises from the fact that the limit equation with $\alpha=0$ is algebraic, so that the boundary conditions cannot be, in general, satisfied. The loss of boundary conditions in a problem leads, usually, to the occurrence of the boundary layer. \\
In the case where the equation is linear, a WKB approximation can be performed which brings transcendentally small terms in the form $\exp(-g(r)/\alpha^n)$, which shows why a simple series expansion $R=\sum_n \alpha^n R_n$ cannot be a correct global approach to the real solution. In fact we see that in a small region near $g(r)\simeq 0$, the terms of this form $\exp(-g(r)/\alpha^n)$ can't be neglected, that small region is called a boundary layer, where the regular expansion fails.\\
Let us consider in this section that $(m,P,\rho)$ are external fields. So, in the limit $\alpha\rightarrow 0$, we can write (\ref{b9:3})  in the form
\begin{align}
\label{eq:ziusudra}
\alpha R''-2 \alpha^2 R'^2-\alpha f(r) R R'-\alpha g(r) R'-h(r) R=k(r)\,,
\end{align}
where $(f,g,h,k)$ are functions of $r$. There are in fact functions of $(m,P,\rho)$.\\
We assume that in some subset of $[0,r_s]$, where $r_s$ is the radius of the star, the solution has a regular expansion of the form
\begin{align}
\label{eq:series}
R(r)=\sum_n \alpha^n R_n(r)\,.
\end{align}
The substitution in the previous equation gives an algebraic equations at each order, 
\begin{align}
R_0=-\frac{k(r)}{h(r)}\,,~~~~R_1=\frac{R_0''-f(r)R_0 R_0'-g(r)R_0'}{h(r)}\,, \cdots
\end{align}
All the coefficients $R_n$ are determined by the previous coefficients, so that we cannot impose the boundary conditions to the regular expansion. Hence the subset where the solution has a regular expansion should not contain the boundary points $r=0$ and $r=r_s$. This part of the solution is known as the outer expansion.\\
We notice that $R_0$ corresponds to the solution in GR and by construction (of the EoS) we know that we have necessarily in GR, $R'(0)=0$, which implies that this condition will also be satisfied for the regular expansion. Therefore the boundary layer exists only near the surface, where the solution will have a fast variation in order to satisfy the conditions $(R(r_s)=R'(r_s)=0)$. Notice that it might be seen as a generalization to curved space-time to the well-known chameleon mechanism.\\
In summary, we have inside the star, a solution very close to GR where we can perform a regular expansion of the form \eqref{eq:series}. Near the surface we use the subtraction trick to determine its nature. We define $C=R-\sum_n \alpha^n R_n$, and near $r=r_s$, we introduce the stretching variable $\xi=(r_s-r) \alpha^{-\mu}$, $\mu>0$, which magnifies the layer. We now assume that there exists a regular expansion for $C(\xi)$, which gives at the lowest order of the expansion near the surface
\begin{align}
\alpha^{1-2\mu}\frac{{\rm} d^2 C_0}{{\rm}d\xi^2}-h(r_s) C_0=0\,,
\end{align}
hence we deduce that $\mu=1/2$ and $C_0=A e^{-\sqrt{h(r_s)}\xi}+B e^{\sqrt{h(r_s)}\xi}$. We can proceed for higher orders by the regular expansion $C=\sum_{n}\alpha^{n/2}C_n$. This solution is the inner expansion which should be matched with the outer solution, hence we fix $B=0$. Imposing the boundary condition $R(0)=0$, we have
\begin{align}
\label{eq:nergal}
R(r)=-\frac{k(r)}{h(r)}+\frac{k(r_s)}{h(r_s)}~e^{-\sqrt{h(r_s)}(r_s-r)/\sqrt{\alpha}}+O(\sqrt{\alpha})\,.
\end{align}
We see that it is very difficult to satisfy the second condition $R'(r_s)=0$, in fact
\begin{align}
\label{eq:anubis}
R'(r_s)=-\frac{k'(r_s)}{h(r_s)}+\frac{k(r_s)h'(r_s)}{h(r_s)^2}+\frac{k(r_s)}{\sqrt{\alpha h(r_s)}}\,.
\end{align}
The presence of the term $\alpha^{-1/2}$ makes the condition very difficult to satisfy. Notice that if we go to the next order of perturbation, we will have an additional term at the surface for $R'$ in the form $-k(r_s)h'(r_s)/(4h(r_s)^2)$, which doesn't cancel the term $\propto \alpha^{-1/2}$. Therefore we understand that in order to satisfy both conditions $R(r_s)=R'(r_s)=0$, we see from \eqref{eq:anubis} that we need to choose carefully the functions $k,h$, which means the EoS.\\
Notice also from \eqref{eq:nergal} that we have oscillations when $h(r_s)<0$, which is equivalent to $\alpha<0$. In fact from \eqref{eq:ziusudra} at the linear order, we have
\begin{align}
\alpha R''-\alpha g(r) R'-h(r) R=k(r)\,,
\end{align}
Hence $\{h,k\}\rightarrow \{-h,-k\}$ is equivalent to $\alpha\rightarrow -\alpha$. Therefore we have an oscillating mode in the ghost case $(\alpha<0)$. We have also noticed, numerically, that we can kill these oscillations if we reduce the EoS to particular cases  where $T(r_s)=T'(r_s)=0$ and $T$ is the trace of the energy momentum tensor.
 
\section{Numerical Procedure and Results}
\label{0directe}  
  
Here in this section, we will discuss about the numerical approach we had undertaken. As a first step, it is convenient to rescale the various quantities involved so as to work directly with dimensionless quantities. We introduced the following rescaled variables :
  
\begin{align}
\label{d1}
r = x\xi_{\star}, \quad m = \bar{m}M_{\odot}, \quad P = \bar{P}P_{\star}, \quad \rho = \bar{\rho}\rho_{\star}\,, \nonumber \\
R = \bar{R}R_{\star}, \quad R'=\bar{R}'\frac{R_{\star}}{\xi_{\star}}, \quad R''=\bar{R}''\frac{R_{\star}}{\xi_{\star}^{2}}, \quad \alpha = \bar{\alpha}\frac{1}{R_{\star}}\,,
\end{align}
where the barred quantities are dimensionless. Substituting the above rescaling \eqref{d1}, we write the set of equations (\ref{b9:1},\ref{b9:2},\ref{b9:3}) in proper dimensions:
 
\begin{widetext}   
\begin{align}
\label{d2:1}
\bar{m}' &= \frac{1}{12c^{2}GM_{\odot}(1+2\bar{\alpha} \bar{R})}\left[\xi_{\star}^{3}x^{2}(48\pi GP_{\star}\bar{P}+c^{4}R_{\star}\bar{R}(2+3\bar{\alpha} \bar{R})+32\pi c^{2}G\rho_{\star}\bar{\rho})\right. \nonumber \\
  &\qquad\qquad\qquad\left.+\frac{3\bar{\alpha}\bar{R}'(-4c^{2}GM_{\odot}\bar{m}(1+2\bar{\alpha} \bar{R}+4\bar{\alpha} x\bar{R}')+\xi_{\star}x^{2}(-16\pi G P_{\star}\xi_{\star}^{2}x\bar{P}+\bar{\alpha} c^{4}(R_{\star}\xi_{\star}^{2}x\bar{R}^{2}+8\bar{R}')))}{1+2\bar{\alpha} \bar{R}+\bar{\alpha} x\bar{R}'}\right]\,, \\
  \label{d2:2}
  \bar{P}' &= \frac{1}{4c^{2}P_{\star}x(c^{2}\xi_{\star}x-2GM_{\odot}\bar{m})(1+2\bar{\alpha}\bar{R}+\bar{\alpha} x\bar{R}')}\left[(P_{\star}\bar{P}+c^{2}\rho_{\star}\bar{\rho})(-4c^{2}GM_{\odot}\bar{m}(1+2\bar{\alpha} \bar{R}\right. \nonumber \\
  &\qquad\qquad\qquad \left.+4\bar{\alpha} x\bar{R}')+\xi_{\star}x^{2}(-16\pi GP_{\star}\xi_{\star}^{2}x\bar{P}+\bar{\alpha} c^{4}(R_{\star}\xi_{\star}^{2}x\bar{R}^{2}+8\bar{R}')))\right]\,,\\
  \label{d2:3}
  \bar{R}'' &= \frac{1}{6\bar{\alpha} c^{2}x(c^{2}\xi_{\star}x-2GM_{\odot}\bar{m})(1+2\bar{\alpha} \bar{R})}\left[\xi_{\star}^{3}x^{2}(1+2\bar{\alpha} \bar{R})(24\pi GP_{\star}\bar{P}+c^{4}R_{\star}\bar{R}-8\pi c^{2} G\rho_{\star}\bar{\rho}) \right. \nonumber \\
  & \qquad\qquad\qquad+ \bar{\alpha} c^{2}(12GM_{\odot}\bar{m}(1+2\bar{\alpha} \bar{R})+c^{2}\xi_{\star}x(-12+\bar{R}(-24\bar{\alpha}+R_{\star}\xi_{\star}^{2}x^{2}\bar{R}))+16\pi G\rho_{\star}\xi_{\star}^{3}x^{3}\bar{\rho})\bar{R}'\nonumber \\
  &\qquad\qquad\qquad\qquad \left.  +12\bar{\alpha}^{2} c^{2}x(c^{2}\xi_{\star}x-2GM_{\odot}\bar{m})\bar{R}'^{2}\right]\,.
  \end{align}
\end{widetext}
Finally, we need to fix the boundary conditions for this system of differential equations. The set of equations (\ref{d2:1},\ref{d2:2},\ref{d2:3}) includes two first-order differential equations and one second-order differential equation. Thus at least four boundary conditions are required to solve the system completely. In order to obtain physically realistic solutions, these boundary conditions must be chosen from the set of  regularity conditions and matching conditions which we give below:
\begin{enumerate}
\item The regularity conditions at the centre of the star demands that the metric functions and the thermodynamic quantities are at least $C^2$ functions of the rescaled radial co-ordinate $x$. Also from the form of the interior metric it is clear that the rescaled ``mass function" $\bar{m}(x)$ should be zero at the centre. Hence at the centre of the star we must have 
\begin{align}
   \label{d3}
    \bar{m}(0) = 0, \quad  
    \bar{P}'(0) = 0, \quad
    \bar{\rho}'(0) = 0, \quad
    \bar{R}'(0) =0
  \end{align}

\item   As we discussed in the earlier section, Schwarzschild solution is the natural exterior solution for this gravity model. To match with a Schwarzschild metric we must have the 
  Ricci scalar and its normal derivative to vanish at the surface. These conditions, along with the matching of the second fundamental form, makes the fluid pressure to vanish at the 
  surface of the star. Hence we have the following boundary conditions on the matching surface $x=x_s$
  \begin{align}
   \label{d4}
    \bar{R}(x_{s}) = 0, \qquad
    \bar{R}'(x_{s}) = 0,  \qquad
    \bar{P}(x_{s})=0.
  \end{align}
 \end{enumerate}
 Any four conditions from the above sets of conditions will in principle solve the system of differential equations. However one has to choose the boundary conditions carefully to avoid any ``unphysical" solution where the energy condition and/or the regularity conditions are violated at any point in the interior of the star.

\section{Direct approach}
\label{directe}
\begin{figure}
\includegraphics[scale=.65]{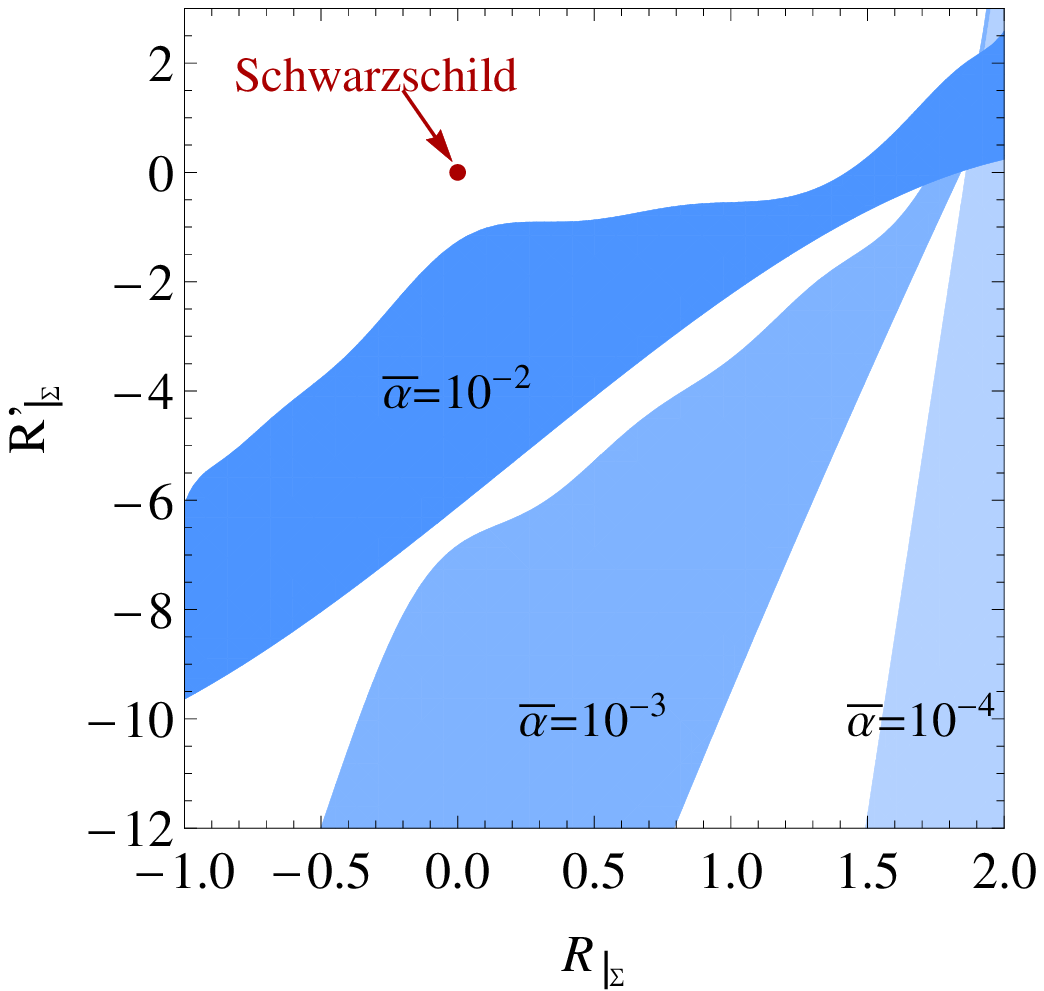}
\includegraphics[scale=.65]{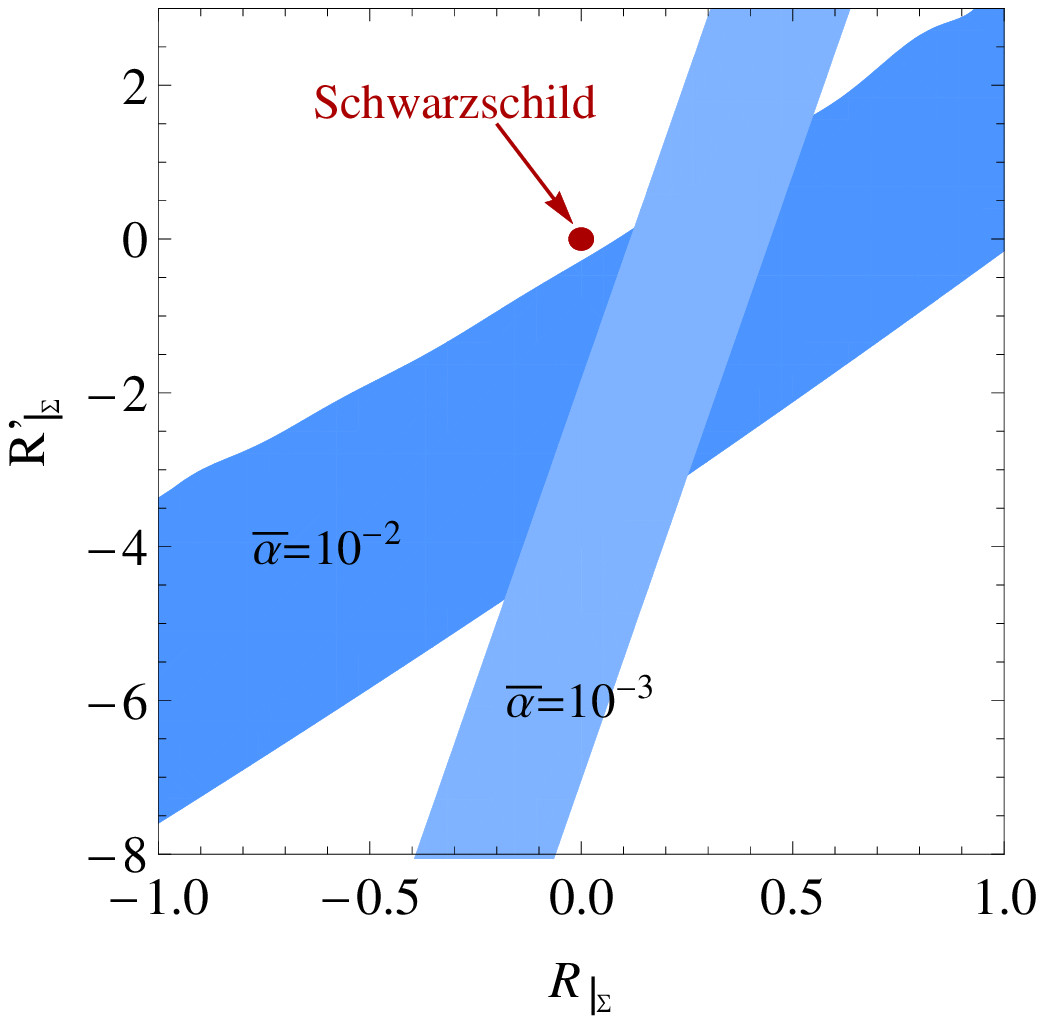}
\caption{On the left panel, the contour plot of $(R,R')$ at the surface of the star ($\Sigma$) for $\bar \rho=1$ and we have considered a large range for the initial conditions $10^{-3}<P(0)<1/3$ and $R(0)>0$. On the right panel, the contour plot of $(R,R')$ at the surface for the polytropic equation $\bar \rho=\bar P^{5/9}/0.2$. In both cases, there are no initial condition which match Schwarzschild at the surface. For smaller $\bar \alpha$ the model goes increasingly far from Schwarzschild in accordance with eq.(\ref{eq:anubis}).}
\label{fig:excluded1}
\end{figure}

\begin{figure}
\begin{center}
\includegraphics[scale=.65]{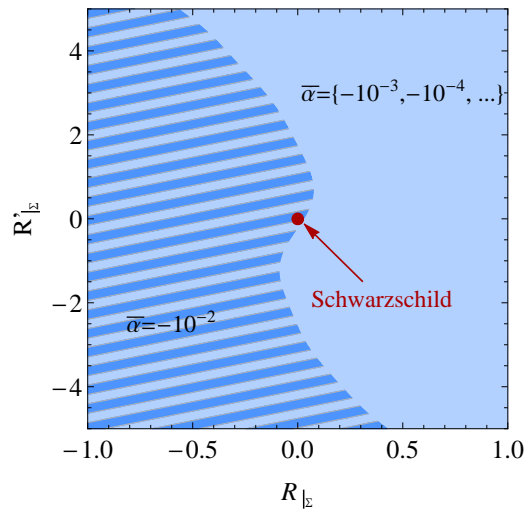}
\end{center}
\caption{Contour plot for $(R,R')$ at the surface ($\Sigma$) of the star for the polytropic equation of state $\bar\rho=\bar P^{5/9}/0.2$ and $\bar \alpha<0$. For a given parameter $\bar \alpha$, there will be a unique solution which matches \textit{exactly} Schwarzschild at the surface. The model is extremely fine-tuned.}
\label{fig:excluded2}
\end{figure}

In this section, we solve directly the set of coupled equations (\ref{d2:1},\ref{d2:2},\ref{d2:3}) along with the boundary conditions (\ref{d3},\ref{d4}). First we have assumed the simplest equation of state $\bar \rho=1$. We have solved the coupled equations for various values of $\bar \alpha$ (see Fig.\ref{fig:excluded1}). We solve the equations by considering random values of $\bar P(0)$ and $\bar R(0)$. Hence the solution gives a star from which we can read the curvature at the surface $(\Sigma)$ $\{\bar R_{|\Sigma},\bar R'_{|\Sigma}\}$. We have considered $10^{-3}<P(0)<1/3$. The lower bound corresponds to a realistic choice of the central pressure, in fact we cannot consider realistic neutron stars with a very small central pressure and the upper bound corresponds to the condition $\rho-3P>0$, always true for relativistic and non-relativistic matter. We have also noticed that if we enlarge the bounds, it will not affect the results. As we can see from Fig.(\ref{fig:excluded2}), we do not match Schwarzschild at the surface. Hence this equation of state is excluded. It is important to notice that this behavior is completely different from General Relativity for which any equation of state is mathematically satisfactory and can be excluded only because of physical reasons. It is also important to notice the evolution of the system when $\bar  \alpha$ decreases, in fact for smaller $\bar \alpha$ the model goes increasingly far from the right boundary conditions at the surface (Schwarzschild). This can be understood from the section \ref{singular}. In fact for smaller $\bar \alpha$, the system will develop a layer bound and hence increasing the difficulty to have Schwarzschild at the surface. To satisfy two conditions at the surface namely $\bar R=0$ and $\bar R'=0$ is impossible for some equation of states.\\
We have also performed the same analysis for a more realistic equation of state (polytropic), $\bar \rho=k \bar P^{5/9}$. We found the same results, we cannot match Schwarzschild at the surface and the star is increasingly far from Schwarzschild at the surface when $\bar \alpha$ decreases. Hence we can conclude that the additional Junction conditions at the surface of the star provides a constraint on the equation of state inside the compact object. Therefore the equation of state should be fine-tuned.\\
Also we found that for the polytropic case, we can match with Schwarzschild if $\bar \alpha<0$, which is obviously excluded because of the ghost condition. But we see that even if an equation of state for $\bar \alpha>0$ is found, we will have to extremely fine-tune the initial conditions $(\bar P(0),\bar R(0))$ in order to satisfy the junction conditions. In fact, only a very particular initial conditions will match our exterior solution. All the other initial conditions will not be correct. Therefore we understand that the model gives rise to two fine-tuning problems. First, only a class of equation of state can be considered (which might be not consistent with particle physics) and second, the initial conditions should be extremely fine-tuned in order to match \textit{exactly} the exterior solution.

\section{Semi analytic approach}
\label{pasdirecte}
As we have seen previously, the exact matching between the interior solution and the exterior solution is sometimes impossible. We have shown the existence of two problems; first only a class of equations of state can be mathematically matched with Schwarzschild, which can be seen as a fine-tuning on the equation of state, and we have also a fine-tuning on the initial conditions, because even for a mathematically viable equation of state, only a particular central condition on the pressure and the curvature will match Schwarzschild at the surface. In this section, we propose a solution to circumvent these difficulties. To avoid these problems, we choose a generic form of the curvature function $ \bar{R}(x)$ that satisfy all the regularity and boundary conditions. Hence the problem on the boundary conditions will be reduced. Also we do not choose a particular equation of state in order to solve the first problem. The equation of state will be determined by the dynamics of the fields. The only boundary condition to satisfy is the pressure at the surface.\\
The most general form of the first derivative of the curvature scalar satisfying these conditions should be
\begin{align}
\label{d5}
\bar{R}'(x)=g(x)x(x-x_{s})\,,
\end{align}
where $g(x)$ is an arbitrary and well defined function of $x$ and $x_s$ is the surface of the star. Integrating $\bar R'$ we can fix the integration constant in order to have the last condition $\bar R(r_s)=0$. Notice that, as in GR, we will always consider $R(r)>0$.\\ 
We now choose a suitable ansatz for the function $g(x)$. By doing so \eqref{d2:3} becomes an algebraic constraint between the pressure and density of the stellar fluid. We also note that by choosing an ansatz for $g(x)$, we can no more specify the equation of state of the stellar matter without over specifying the system. Hence the equation of state will be determined by the solution of the system. If a certain class of $g(x)$ gives unphysical equation of state, we will discard the class. As the simplest choice of the ansatz,  let us consider the generic function $g(x)$ to be a constant. But the system will not satisfy an additional condition, namely $R''(r_s)\le 0$. Indeed from eq.(\ref{b9:3}) and the conditions at the surface, $R=R'=P=0$, we have
\begin{align}
  R'' =- \frac{4\pi r_s }{3\alpha [r_s-2m(r_s)]} \rho \le 0.
\end{align}
Therefore we will assume $g(x)=A(x-x_s)$ which satisfies all conditions  and solve the dynamical equations to get the mass and pressure profiles of a neutron star and its corresponding EoS.  Hence the curvature scalar has the form 

\begin{figure}[h]
\includegraphics[scale=0.6]{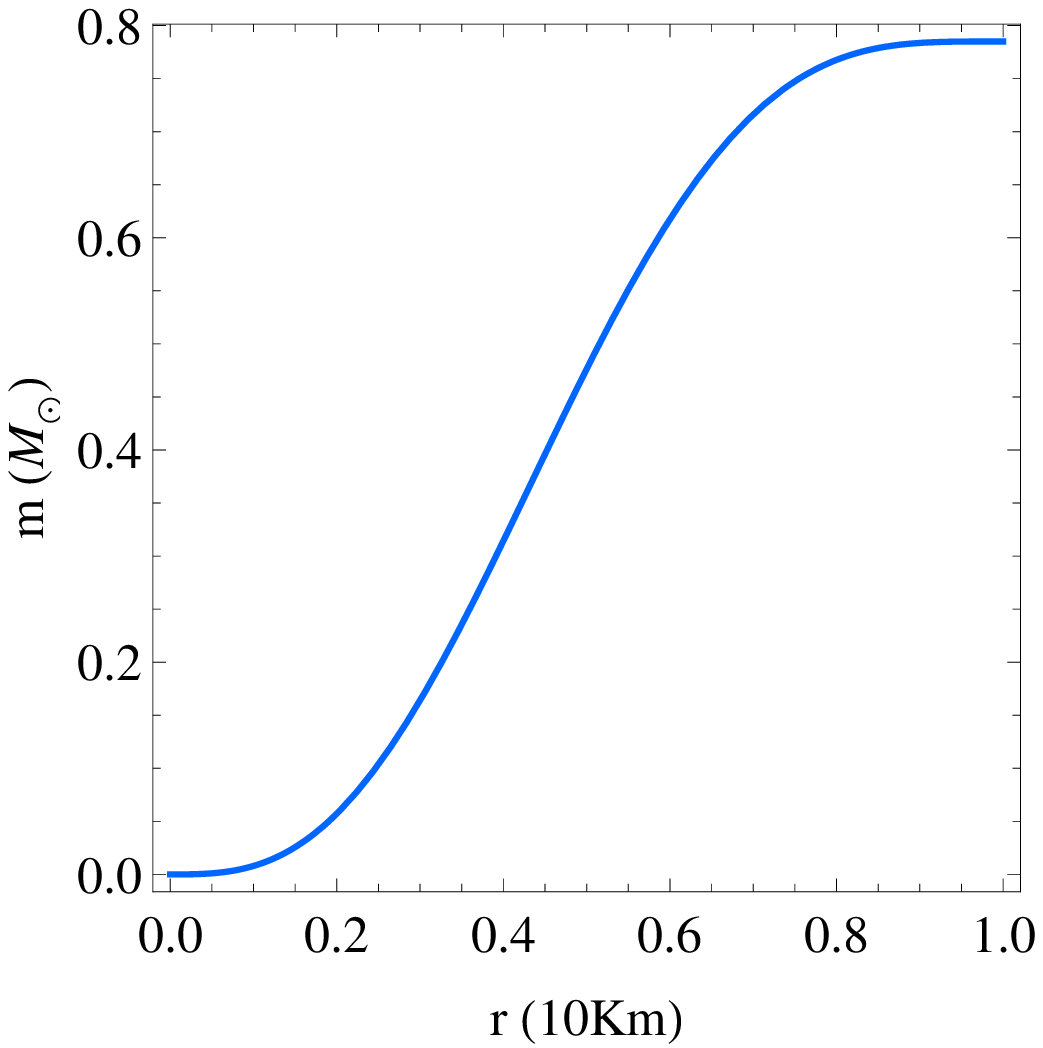} 
\includegraphics[scale=0.6]{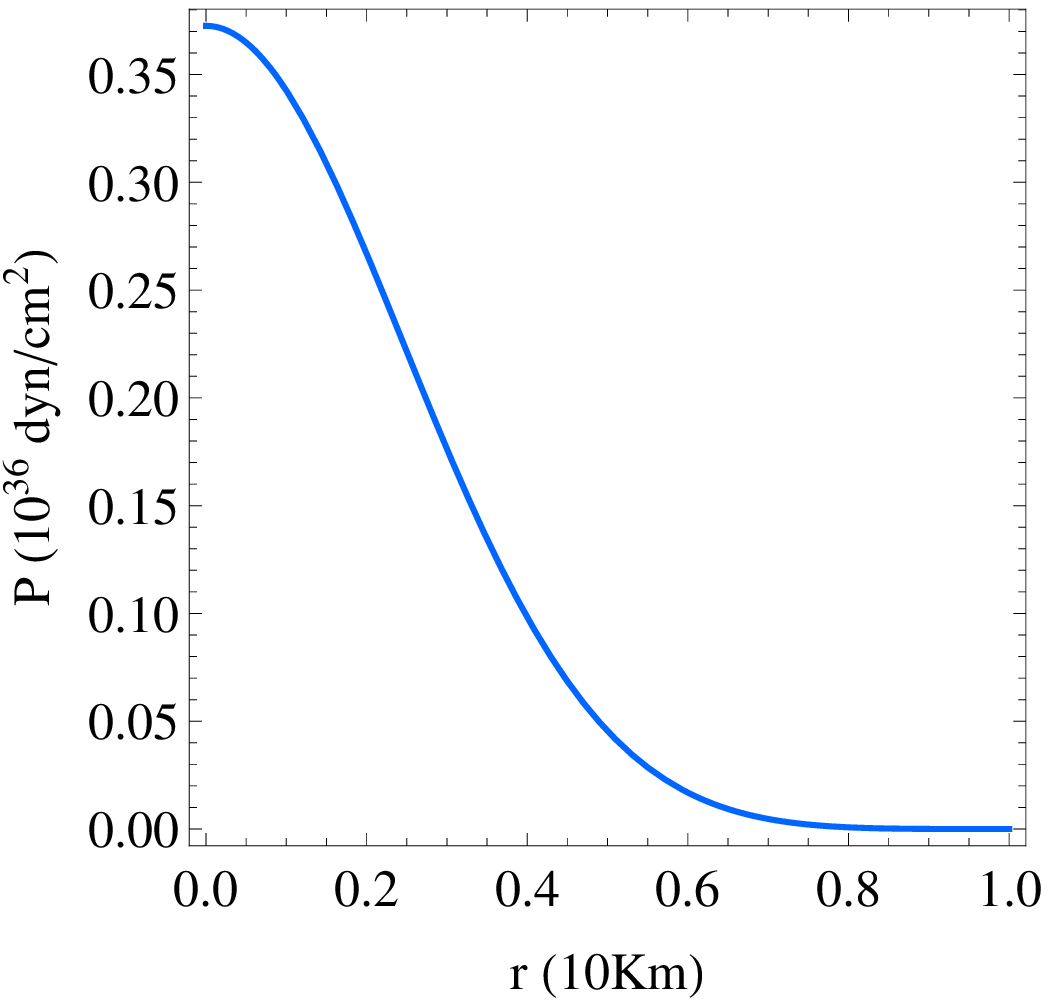} 
\includegraphics[scale=0.6]{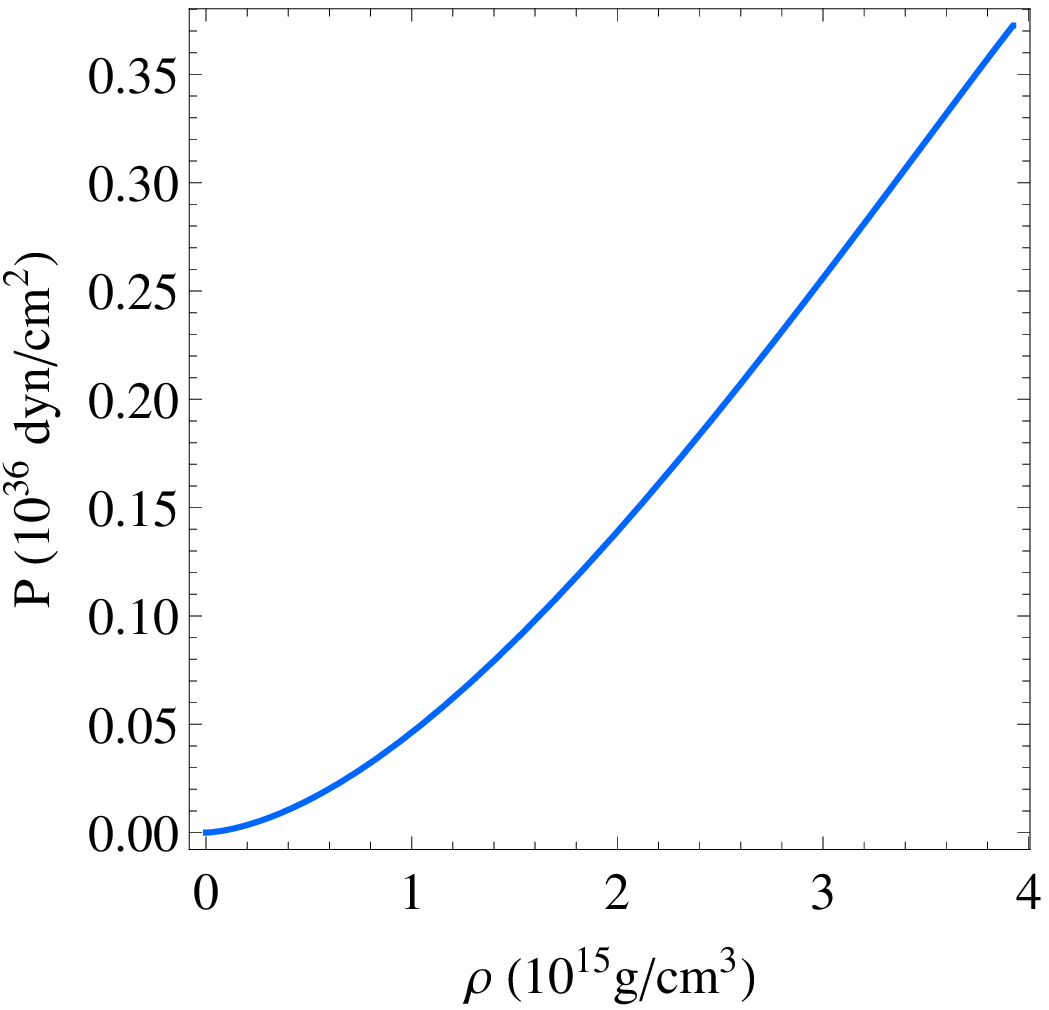}
\caption{Mass, Pressure and EoS profiles for $\bar{R}=\frac{A}{12}(x-1)^{3}(3x+1)$, $A=-60$ and $\alpha=10^{-45}$ eV$^{-2}$.}
\label{fig:pasdirecte}
\end{figure}

\begin{align}
\label{d6}
\bar{R}(x)=\frac{A}{12}(x-x_{s})^{3}(3x+x_{s})
\end{align}
which satisfies the boundary conditions. The mass and pressure profiles of the star and the corresponding EoS for the given form of $\bar{R}$ are shown in Fig.(\ref{fig:pasdirecte}). We have taken $x_{s}=1$ i.e. fixed the radius of the star to 10 Km and the central pressure is chosen in order to have $\bar P(x_s)=0$ at the fixed position of the surface $x_s=1$.\\
So the mass of the star comes out to be $0.79~M_{\odot}$ and its central pressure is 3.73$\times 10^{35}$ dynes/cm$^{2}$. Fitting the EoS, we get a profile of the form 
\begin{align}
\label{d7}
\bar{P}\simeq 0.049~\bar{\rho}^{1.49}
\end{align}
Therefore we have shown in this section that it is possible to match exactly Schwarzschild at the surface of the star for realistic parameters (e.g. $\alpha=10^{-45}$ eV$^{-2}$) and the star obtained is perfectly realistic.

\section{Conclusion}
In this work, we have derived for the first time the full exact solution for a neutron star in Starobinsky model with the exterior matching solution, namely Schwarzschild. However in this model, and also in other $f(R)$ modified gravity models in general, the field equations are highly non-linear. Difficulty crops in due to the fourth order of the field equations hence on the necessity to satisfy extra two junction conditions, namely the continuity of the Ricci scalar and its normal derivative. This makes the problem more stringent and has not been considered before for studying compact stars in these models. Also, for the $R+\alpha R^{2}$ model that we have considered in this paper, Schwarzschild is the only static spherically symmetric asymptotically flat solution with a regular horizon which forces us to match star with a Schwarzschild exterior.  We have shown that the equations are singular in the sense that they develop a boundary layer, hence all boundary conditions cannot be satisfied for a generic EoS. Only a particular class of EoS will be compatible with the model. While trying to solve the system in a direct numerical approach, we face the typical singularity problem in dealing with the small value of the coupling constant $\alpha$. But even for higher values of $\alpha$, choosing constant density and polytropic EoS, we found that matching with Schwarzschild exterior is not possible. In fact, we have also shown that as we go to smaller values of $\alpha$, the system moves further away from Schwarzschild. This clearly indicates that in this model some EoS can be ruled out even without considering observational constraints unlike General Relativity. Also, for negative $\alpha$, which has its inherent ghost problem, we found the Schwarzschild exterior solution only for certain fine-tuned initial conditions. Therefore, the model brings two additional fine-tuning problems. First only a class of equations of state can be mathematically matched with Schwarzschild and second, the central initial conditions should be fine-tuned in order to match exactly Schwarzschild at the surface. From our point of view if we assume that the equation of state is fixed and can't fluctuate, hence only a very particular set of boundary conditions will produce the star, an extremely small deviation from that set of conditions will not be a solution, which implies an instability of a solution. 
\\
So, the obvious question which arises is whether there exist any solution in this model which smoothly matches with Schwarzschild or not. To check that, we took a semi-analytical approach to choose the form of the Ricci scalar and its first derivative which satisfies all the boundary conditions. We solved the resultant system to get the mass and the pressure profile which is physically viable and also an acceptable EoS which proves that there exist a class of solution, which apart from satisfying all the boundary and junction conditions, matches smoothly with Schwarzschild exterior solution. \\
Therefore we have shown that the matching with Schwarzschild at the surface is possible but highly unnatural. And this phenomenon is true for a wide class of $f(R)$ theories that 
permit the Schwarzschild spacetime and also satisfy the uniqueness theorem. This is unnatural because the EoS of the compact star matter should be completely determined by the nuclear physics and the macroscopic description of quantum field theory for a highly densed star and not by the theory of gravity. \\
We should emphasize that the matching with any exterior Ricci flat solution will bring the same difficulties, because of the boundary conditions $R=R'=0$ at the surface of the star, and this result can be simply generalized for any viable $f(R)$. Therefore modelling a radiating star with an intermediate Vaidya region is equally fine tuned and unnatural.\\ 
At this point one may argue that in realistic astrophysical scenarios the domain of applicability of the spherically symmetric and vacuum conditions are typically set at astrophysical scales. For example in case of our solar system the domain is within the heliopause which is approximately one light year from the sun. Hence the solar system is not in the 
{\it ideal} sense asymptotically flat. However within this domain the solar system definitely is "almost" spherically symmetric and "almost" vacuum (with respect to the scales of the problem) and hence Schwarzschild solution is a very good approximation.  Even in the case of $f(R)$ gravity, the conditions on validity and stability of the Jebsen-Birkhoff theorem, ensure that the solution will remain ``almost" Schwarzschild in the exterior domain for Starobinsky model provided the value of $\alpha$ is small \cite{nzioki}. Unfortunately the instability of our solution that is described in section VI, arising due to the matching of $R' $ at the boundary for very small $\alpha$ still remains. \\
We would like to emphasize here that our study should not be seen as an exact realization of a real situation but more as a limit of various physical and realistic problems. In all cases, either we have rotation or matter in the exterior, we should keep in mind that in the limit where the spinning goes to zero or the limit where the density of matter goes to zero, our result should be recovered. Assuming that physics is not discontinuous, we can think that our problem is a good approximation of a real situation.\\
A possible solution will be the matching with an exterior solution of the following form:
\begin{align}
m(r)=M-a e^{-br}
\end{align}
This solution is asymptotically Schwarzschild, hence it can match with the standard constraints in the solar system provided $(b)$ is large enough. For this solution, we have 
\begin{align}
\label{eq:end}
R=2ab(br-2)\frac{e^{-b r}}{r^2}
\end{align}
Hence we will be able to alleviate the fine-tuning problem. In fact, any numerical solution can be matched at the surface with (\ref{eq:end}) by choosing different values of $(a,b)$.\\
Even if that class of solutions exists for other $f(R)$ models, it is well-known that for the Starobinsky model, Schwarzschild is the unique asymptotically flat solution. Hence the fine-tuning problem cannot be circumvented in that case.\\
Another way to address this fine tuning problem would be to neglect the extra matching conditions at the surface (namely the matching of Ricci scalar and it's normal derivative)  and let there be a delta function in the field equations, which will give rise to a surface stress-energy term. This procedure was used to model static gravstars in General Relativity \cite{Mazur}.
However this will radically change the structure of the crust of the neutron stars and should have, in principle, observational signatures \cite{Lattimer}.

\section*{Acknowledgments}
  
AG, RG1 and RG2 want to thank the National Research Foundation and the University of KwaZulu-Natal for financial support.  SR acknowledges the NRF incentive grant for research support.


\begin{thebibliography}{99}


\bibitem{Guth:1980zm}
  A.~H.~Guth,
  Phys.\ Rev.\ D {\bf 23} (1981) 347.

\bibitem{Linde:1981mu}
  A.~D.~Linde,
  Phys.\ Lett.\ B {\bf 108} (1982) 389.

\bibitem{Albrecht:1982wi}
  A.~Albrecht and P.~J.~Steinhardt,
  Phys.\ Rev.\ Lett.\  {\bf 48} (1982) 1220.

\bibitem{Mukhanov:1981xt}
  V.~F.~Mukhanov and G.~V.~Chibisov,
  JETP Lett.\  {\bf 33} (1981) 532
   [Pisma Zh.\ Eksp.\ Teor.\ Fiz.\  {\bf 33} (1981) 549].

\bibitem{Hawking:1982cz}
  S.~W.~Hawking,
  Phys.\ Lett.\ B {\bf 115} (1982) 295.

\bibitem{Starobinsky:1982ee}
  A.~A.~Starobinsky,
  Phys.\ Lett.\ B {\bf 117} (1982) 175.

\bibitem{Guth:1982ec}
  A.~H.~Guth and S.~Y.~Pi,
  Phys.\ Rev.\ Lett.\  {\bf 49} (1982) 1110.

\bibitem{Ginzburg}
V. I. Ginzburg, D. A. Kirzhnits, \& A. A. Lyubushin, 1971 Zh. eksp. teor. Fiz. 60, 451.
(Sov. Phys. JETP 33, 24271). 
\bibitem{Bunch}
T. S. Bunch, P. C. W. Davies, Stress tensor and conformal anomalies for massless fields in a Robertson-Walker
universe, King's College preprint (1976). 
\bibitem{Davies}
P.~C.~W.~Davies,  Phys.\ Lett.\ B {\bf 68} (1977) 402.

\bibitem{Starobinsky:1980te} 
  A.~A.~Starobinsky,
  Phys.\ Lett.\ B {\bf 91}, 99 (1980).

\bibitem{Thongkool:2009js}
  I.~Thongkool, M.~Sami, R.~Gannouji and S.~Jhingan,
  Phys.\ Rev.\ D {\bf 80} (2009) 043523
  [arXiv:0906.2460 [hep-th]].

\bibitem{Amendola:2006we}
  L.~Amendola, R.~Gannouji, D.~Polarski and S.~Tsujikawa,
  Phys.\ Rev.\ D {\bf 75} (2007) 083504
  [gr-qc/0612180].

\bibitem{Frolov:2008uf}
  A.~V.~Frolov,
  Phys.\ Rev.\ Lett.\  {\bf 101} (2008) 061103
  [arXiv:0803.2500 [astro-ph]].

\bibitem{Kobayashi:2008tq}
  T.~Kobayashi and K.~-i.~Maeda,
  Phys.\ Rev.\ D {\bf 78} (2008) 064019
  [arXiv:0807.2503 [astro-ph]].

\bibitem{Upadhye:2009kt}
  A.~Upadhye and W.~Hu,
  Phys.\ Rev.\ D {\bf 80} (2009) 064002
  [arXiv:0905.4055 [astro-ph.CO]].

\bibitem{Appleby:2008tv}
  S.~A.~Appleby and R.~A.~Battye,
  JCAP {\bf 0805} (2008) 019
  [arXiv:0803.1081 [astro-ph]].

\bibitem{Faraoni:2011pm}
  V.~Faraoni,
  Phys.\ Rev.\ D {\bf 83} (2011) 124044
  [arXiv:1106.0328 [gr-qc]].
      
\bibitem{Ade:2013uln} 
  P.~A.~R.~Ade {\it et al.}  [Planck Collaboration],
  arXiv:1303.5082 [astro-ph.CO].
  
\bibitem{Psaltis:2008bb} 
  D.~Psaltis,
  Living Rev. Relativity, {\bf 11}, 9 (2008);
  [arXiv:0806.1531 [astro-ph]].

\bibitem{Kainulainen:2007bt}
  K.~Kainulainen, J.~Piilonen, V.~Reijonen and D.~Sunhede,
  Phys.\ Rev.\ D {\bf 76} (2007) 024020
  [arXiv:0704.2729 [gr-qc]].

\bibitem{Babichev:2009td}
  E.~Babichev and D.~Langlois,
  Phys.\ Rev.\ D {\bf 80} (2009) 121501
   [Erratum-ibid.\ D {\bf 81} (2010) 069901]
  [arXiv:0904.1382 [gr-qc]].

\bibitem{Cooney:2009rr} 
  A.~Cooney, S.~DeDeo and D.~Psaltis,
  Phys.\ Rev.\ D {\bf 82}, 064033 (2010)
  [arXiv:0910.5480 [astro-ph.HE]].

\bibitem{Babichev:2009fi}
  E.~Babichev and D.~Langlois,
  Phys.\ Rev.\ D {\bf 81} (2010) 124051
  [arXiv:0911.1297 [gr-qc]].
 
\bibitem{Arapoglu:2010rz} 
  A.~S.~Arapoglu, C.~Deliduman and K.~Y.~Eksi,
  JCAP {\bf 1107}, 020 (2011)
  [arXiv:1003.3179 [gr-qc]].
  
\bibitem{Orellana:2013gn} 
  M.~Orellana, F.~Garcia, F.~A.~Teppa Pannia and G.~E.~Romero,
  Gen.\ Relativ.\ Gravit.\ 
  [arXiv:1301.5189 [astro-ph.CO]].

\bibitem{Alavirad:2013paa}
  H.~Alavirad and J.~M.~Weller,
  arXiv:1307.7977 [gr-qc].

\bibitem{Astashenok:2013vza}
  A.~V.~Astashenok, S.~Capozziello and S.~D.~Odintsov,
  arXiv:1309.1978 [gr-qc].

\bibitem{Israel:1966rt} 
  W.~Israel,
  Nuovo Cim.\ B {\bf 44S10}, 1 (1966)
  [Erratum-ibid.\ B {\bf 48}, 463 (1967)]
  [Nuovo Cim.\ B {\bf 44}, 1 (1966)].
  
\bibitem{Deruelle:2007pt} 
  N.~Deruelle, M.~Sasaki and Y.~Sendouda,
  Prog.\ Theor.\ Phys.\  {\bf 119}, 237 (2008)
  [arXiv:0711.1150 [gr-qc]].
  
  \bibitem{clifton}
  T. Clifton, P.K.S. Dunsby, R. Goswami, A. M. Nzioki.
  Phys. Rev. D {\bf 87 }  063517 (2013)
 [arXiv:1210.0730 [gr-qc]].
 
\bibitem{senovilla} 
  J.~ M.~M.~Senovilla,
  arXiv:1303.1408 [gr-qc].
  
  \bibitem{clifton1} 
  T.~Clifton,
  Class.\ Quant.\ Grav.\  {\bf 23}, 7445 (2006)
  [gr-qc/0607096].
  
\bibitem{Nzioki:2009av} 
  A.~M.~Nzioki, S.~Carloni, R.~Goswami and P.~K.~S.~Dunsby,
  Phys.\ Rev.\ D {\bf 81}, 084028 (2010)
  [arXiv:0908.3333 [gr-qc]].

\bibitem{Whitt:1984pd}
  B.~Whitt,
  Phys.\ Lett.\ B {\bf 145} (1984) 176.
           
\bibitem{Mignemi:1991wa} 
  S.~Mignemi and D.~L.~Wiltshire,
  Phys.\ Rev.\ D {\bf 46}, 1475 (1992)
  [hep-th/9202031].

\bibitem{Bekenstein:1995un}
J.~D.~Bekenstein,
Phys.\ Rev.\ D {\bf 51} (1995) 6608.

\bibitem{Sudarsky:1995zg}
D.~Sudarsky,
Class.\ Quant.\ Grav.\ {\bf 12} (1995) 579.

\bibitem{Sotiriou:2011dz}
T.~P.~Sotiriou and V.~Faraoni,
Phys.\ Rev.\ Lett.\ {\bf 108} (2012) 081103
[arXiv:1109.6324 [gr-qc]].

\bibitem{Torii:1998ir}
T.~Torii, K.~Maeda and M.~Narita,
Phys.\ Rev.\ D {\bf 59} (1999) 064027
[gr-qc/9809036].

\bibitem{Starobinsky:2007hu}
  A.~A.~Starobinsky,
  JETP Lett.\  {\bf 86} (2007) 157
  [arXiv:0706.2041 [astro-ph]].
      
\bibitem{Khoury:2003aq}
  J.~Khoury and A.~Weltman,
  Phys.\ Rev.\ Lett.\  {\bf 93} (2004) 171104
  [astro-ph/0309300].

\bibitem{Gannouji:2012iy}
  R.~Gannouji, M.~Sami and I.~Thongkool,
  Phys.\ Lett.\ B {\bf 716} (2012) 255
  [arXiv:1206.3395 [hep-th]].


\bibitem{Stelle:1977ry}
  K.~S.~Stelle,
  Gen.\ Rel.\ Grav.\  {\bf 9} (1978) 353.
  
\bibitem{Kapner:2006si}
  D.~J.~Kapner, T.~S.~Cook, E.~G.~Adelberger, J.~H.~Gundlach, B.~R.~Heckel, C.~D.~Hoyle and H.~E.~Swanson,
  Phys.\ Rev.\ Lett.\  {\bf 98} (2007) 021101
  [hep-ph/0611184].
 
\bibitem{Zhang:2007ne}
  P.~-J.~Zhang,
  Phys.\ Rev.\ D {\bf 76} (2007) 024007
  [astro-ph/0701662].


\bibitem{Starobinsky:1983zz}
  A.~A.~Starobinsky,
  Sov.\ Astron.\ Lett.\  {\bf 9} (1983) 302.
\bibitem{Upadhye:2012vh}
  A.~Upadhye, W.~Hu and J.~Khoury,
  Phys.\ Rev.\ Lett.\  {\bf 109} (2012) 041301
  [arXiv:1204.3906 [hep-ph]].
  
\bibitem{Nayfeh}
Ali H. Nayfeh,
Perturbation Methods,
Wiley-VCH (2000) 437 p;
Mark H. Holmes,Introduction to Perturbation Methods,
Springer (1995) 369 p.


  \bibitem{nzioki} 
  A.~M.~Nzioki, R.~Goswami and P.~K.~S.~Dunsby,
  arXiv:1312.6790 [gr-qc].


\bibitem{Mazur} 
  P.~O.~Mazur and E.~Mottola,
  Proc.\ Nat.\ Acad.\ Sci.\  {\bf 101}, 9545 (2004)
  [gr-qc/0407075].
  
\bibitem{Lattimer} 
  J.~M.~Lattimer and M.~Prakash,
  Astrophys.\ J.\  {\bf 550}, 426 (2001)
  [astro-ph/0002232].
  
  
   
\end{thebibliography}
\end{document}